\documentclass[reprint,amsmath,amssymb,aps,]{revtex4-2}
\usepackage{graphicx}
\usepackage{dcolumn}
\usepackage{bm}
\usepackage{subcaption}
\usepackage{xcolor}
\usepackage{appendix}
\usepackage{float}
\usepackage{titlesec}
\begin{document}

\title{Matter creation, adiabaticity and phantom behavior}
\author{V\'ictor H. C\'ardenas$^1$ }
\author{Samuel Lepe $^2$}
\affiliation{$^1$ Instituto de Fisica y Astronomia, Facultad de Ciencias, Universidad de Valparaíso, Gran Bretaña 1111, Valparaíso, Chile \\ $^2$ Instituto de Física, Facultad de Ciencias, Pontificia Universidad Cat\'olica de Valparaiso, Avenida Brasil 2950, Valparaíso, Chile.}

\date{\today}

\begin{abstract}
We present a novel cosmological framework that unifies matter creation dynamics with thermodynamic principles. Starting with a single-component fluid characterized by a constant equation of state parameter, \(\omega\), we introduce a generalized second law of thermodynamics by considering the entropy associated with the cosmic horizon. Imposing an adiabatic expansion condition uniquely determines the particle creation rate, \(\Gamma\), a feature unprecedented in previous matter creation models. This mechanism yields a cosmology featuring phantom-like expansion while relying solely on a single constituent, which can be either a quintessence-like fluid or a non-exotic, non-relativistic dark matter component. Remarkably, this framework avoids the need for exotic physics while providing a consistent explanation for the accelerated expansion of the universe. Our results open new pathways for understanding the interplay between horizon thermodynamics, particle creation, and cosmic evolution, offering fresh insights into the nature of dark energy and its potential thermodynamic origins.
\end{abstract}

\maketitle

\section{Introduction}

The accelerated expansion of the universe, first discovered through observations of type Ia supernovae \cite{SupernovaSearchTeam:1998fmf}, \cite{SupernovaCosmologyProject:1998vns}, remains one of the most profound mysteries in modern cosmology. This phenomenon, often attributed to a mysterious form of energy known as dark energy \cite{Frieman2008}, \cite{Weinberg2013}, lacks a definitive theoretical explanation despite decades of research. Among the numerous approaches proposed to tackle this enigma, those grounded in fundamental principles from unification theories, such as the holographic principle \cite{tHooft:1993dmi}, \cite{Fischler:1998st}, stand out as particularly compelling.

The holographic principle \cite{Bousso:2002ju} posits that all the information contained within a region of space (the bulk) can be fully described by a theory defined on the boundary of that region, with the degrees of freedom of the boundary theory encoding the bulk dynamics. The first attempts to apply this idea to cosmology was \cite{Fischler:1998st}, and it was followed by several improvements \cite{cohen}, \cite{Hsu:2004ri}, and \cite{li}, which finally led to a working structure that has been used in many successful applications. For example, one such model, entropic cosmology \cite{efs11}, derives cosmic acceleration from the thermodynamic behavior of a holographic screen. This approach offers a stark contrast to the \(\Lambda\)CDM model, which, while successful in describing the large-scale structure and dynamics of the universe, provides an unsatisfactory and often ill-defined thermodynamic description at late times.

However, although successful in describing a cosmological model that recreates the dynamics of the universe, most of these models add new functions and parameters whose origin and meaning are difficult to establish, and of course they manage to fit the data, but because they increase the space of parameters to fit. What we need is to integrate the idea of holography into a model that can describe the accelerated expansion without the need to expand the number of free functions and parameters.

A promising approach to addressing these challenges involves integrating matter creation mechanisms, initially proposed by Prigogine et al. \cite{pggn89} and subsequently developed by Calvão, Lima, and Waga \cite{clw92}. Rooted in nonequilibrium thermodynamics, this mechanism facilitates the creation of particles in an expanding universe, dynamically introducing additional energy density contributions. A key advantage is its reliance on a single component, typically cold dark matter, to describe an accelerating universe without invoking exotic new components. Numerous studies have investigated this model \cite{Qiang2007, Lima2008, Steigman2009, Jesus2011, Lima2012, Cardenas:2012kg, Carneiro2014, Alcaniz2014, Fabris2014, Nunes2016, Pan2016, Jesus2017, CCL2020, Ivanov2019, Trevisani2023}, though most assume an ad-hoc particle creation rate, \(\Gamma\), lacking fundamental justification—effectively substituting one unknown (\(\Lambda\)) with another (\(\Gamma\)). A novel model linking the matter creation process to holography was first explored in \cite{ccl22}, demonstrating the potential of such mechanisms to alter cosmological dynamics in ways consistent with observational data. Furthermore, \cite{CC2024} derived and analyzed a general form of \(\Gamma\) capable of replicating any dark energy model.

In this paper, we integrate the second law of thermodynamics by considering both the entropy generated through matter creation and the entropy associated with the horizon. Our findings show that incorporating matter creation alters the effective thermodynamic pressure and energy density of the cosmic fluid, leading to new dynamics that address certain criticisms of the original entropic cosmology model. This approach also offers a foundation for investigating the interplay between entropy production, nonequilibrium thermodynamics, and cosmological evolution. Throughout this work, we adopt the unit system \(8\pi G = c = k_B = 1\).

\section{Matter creation scenario}

Let us assume that the particle number $N^{\mu}=nu^{\mu}$ is not conserved $N^{\mu}_{;\mu}=n\Gamma$, where $n$ is the number density, $u^{\mu}$ is the 4-velocity, and $\Gamma$ is the particle creation rate. From this is clear that
\begin{equation}\label{ene}
  \dot{n} + 3 H n = n\Gamma,
\end{equation}
for the non conserved number density $n$, where $H=\dot{a}/a$ is the Hubble function and $a$ the scale factor. Also, the vector entropy flux $S^{\mu}=\sigma N^{\mu}=\sigma n u^{\mu}$ where $\sigma$ is the specific entropy, satisfy the second law
\begin{equation*}
    S^{\mu}_{; \mu}=n\dot{\sigma} + n\sigma \Gamma \geq 0.
\end{equation*}
Considering a single component, from
\begin{equation}\label{eq: firstlaw}
    TdS = dU +PdV -\mu dN,
\end{equation}
where is clear $S=S(U,V,N)$. Writing this equation in terms of the densities $\rho=U/V$, $s=S/V$ and $n=N/V$ we get
\begin{equation}
    (Ts-P-\rho +\mu n)dV + (Tds-d\rho +\mu dn)V =0,
\end{equation}
from which we get
\begin{equation}
    s = \frac{P+\rho -\mu n}{T}.
\end{equation}
Now, using the notation $s = n \sigma $ we can write
\begin{equation}\label{eq: mueq}
    \mu = \frac{\rho +P}{n} - T\sigma.
\end{equation}
Assuming that $\dot{\sigma} =0$ we can combine (\ref{eq: firstlaw}) and (\ref{eq: mueq}) to get
\begin{equation}\label{eq: firstlaw2}
    d(\rho V) + PdV = \frac{\rho + P}{n} d(nV).
\end{equation}
It is important to stress that the universe evolution continues to be adiabatic, in the sense that the entropy per particle $\sigma$ remains unchanged. Now, from (\ref{eq: firstlaw2}) we can write it in the usual form of a conserved equation, with now having an extra pressure term
\begin{equation}\label{eq: drhoV}
    d(\rho V) = -(p + p_c)dV,
\end{equation}
where we have identified
\begin{equation}\label{eq: pc1}
    p_c = -\frac{\rho +P}{n}\frac{d(n V)}{dV}.
\end{equation}
This expression can be rewritten in terms of $\Gamma$ using Eq.(\ref{ene}) 
\begin{equation}\label{pc}
p_c=-\left(\frac{\rho + P}{3H}\right) \Gamma.
\end{equation}
This is the source that produces the acceleration of the
universe expansion. Once the particle number increases with the volume, we obtain a negative pressure.

Considering creation of matter implies we have to consider the second law constraint for this contribution
\begin{equation}\label{second}
dS=\frac{s}{n}d(nV) \geq 0,
\end{equation}
where $s=S/V$ is the entropy density. From (\ref{eq: firstlaw}) we find that the new set of Einstein equations are
\begin{equation}\label{friedman}
  3H^2 =  \rho
\end{equation}
\begin{equation}\label{rho}
  \dot{\rho} = \frac{\dot{n}}{n}(\rho + p)
\end{equation}
and the previously derived relation (\ref{ene}). The set of
equations (\ref{ene}), (\ref{friedman}) and (\ref{rho}) completely
specified the system evolution. It is also useful to combine
(\ref{pc}) with (\ref{friedman}) and (\ref{rho}) to eliminate $\rho$
and obtain
\begin{equation}\label{eq5}
2\frac{\ddot{a}}{a}+\frac{\dot{a}^2}{a^2}= -(p +
p_c),
\end{equation}
or in terms of the deceleration parameter $q = -1 - \dot{H}/H^2$ we can write
\begin{equation}
    q = \frac{1}{2} \left(1 + \frac{3}{\rho} (p + p_c) \right).
\end{equation}
The standard adiabatic evolution is easily recovered: setting
$\Gamma = 0$ implies that $\dot{n}/n=-3H$, which leads to the usual
conservation equation from (\ref{rho}). A class of de Sitter
solution is obtained with $\dot{n}=\dot{\rho}=0$ and arbitrary
pressure $p$. Moreover, there exist another class of solutions where
Eq.(\ref{rho}) enable us to determine the pressure; for example if
$\rho=m n$, with $m$ constant, Eq.(\ref{rho}) implies $p=0$, and
furthermore if $\rho=aT^4$ and $n=bT^3$ implies $p=\rho/3$.

\section{Incorporating holographic ideas }

In this section, we propose a new venue that enables us to fit the data and give a more complete understanding of the thermodynamics of the late-time universe evolution. The idea is based on considering the holographic contribution from the horizon together with the effect of particle creation first considered by Prigogine \cite{pggn89}. 

From a thermodynamical point of view, this is a natural way to proceed. According to Prigogine, there are two types of entropies: $dS_e$ associated to ``entropy flow'' and the irreversible one $dS_i$ associated to ``entropy creation'', where $dS_i \geq 0$. It is natural to associate $dS_e$ to the entropy gradient of the horizon. This change in entropy is due to the expansion of the universe. However, $dS_i$ is related to the creation of particles with rate $\Gamma$. In this case, the total entropy change is $dS = dS_e+dS_i$ as in Prigogine work. 

Considering the entropy associated to the horizon, we get
\begin{equation*}
    S = \frac{A}{4} = \frac{\pi}{H^2},
\end{equation*}
where $A=4\pi r_h^2$ and we have used that $r_h = H^{-1}$. In this case the growth of the entropy follows
\begin{equation} \label{eq: dse}
    dS_{e} = - 2 \pi \frac{dH}{H^3}.
\end{equation}
On the other hand, the creation of particles generates a term $dS_i$ which is related to $\Gamma$, the rate of particle creation. Based on (\ref{ene}) and (\ref{second}) in this case
\begin{equation}\label{eq: dis}
    dS_i = n\sigma  \frac{\Gamma}{3H}dV.
\end{equation}

Assuming that these effects are connected through the concept of an expanding adiabatic universe, where the increase in horizon entropy is compensated by the entropy generated through particle production, allows for the determination of the particle production rate. The next question is whether this specific form of \(\Gamma\) results in a viable cosmological model, which will be explored in the following.

Let us write the previous results in a more transparent way. Introducing the deceleration parameter $q=-1-\dot{H}/H^{2}$ we can write (\ref{eq: dse}) as
\begin{equation}
dS_{e}=-\frac{2\pi }{H^2}\left( 1+q\right) \frac{dz}{1+z},  \label{A5}
\end{equation}
where we have used that $a = (1+z)^{-1}$. Now, considering the particle creation contribution, we get from (\ref{eq: dis}) that
\begin{equation}\label{eq: A6}
dS_{i} = - \frac{\sigma N \Gamma}{H} \frac{dz}{1+z} , 
\end{equation}%
where we have used that $dS_i = \sigma d(n V)$ with $V=a^3$ and by imposing adiabaticity $dS=dS_e+dS_i=0$, we get
\begin{equation}
\Gamma =-\frac{2\pi }{\sigma N H}\left( 1+q\right).
\label{A9}
\end{equation}%
From this relation we can get a positive $\Gamma$, i. e. a particle creation phase, only if we have $q<-1$, typically of a phantom-dominated regime. The other case, with $q>-1$ implies $\Gamma <0$ indicating a phase of particle annihilation. 

Let us consider an explicit EoS $p=\omega \rho$ for the matter content. In this case, from (\ref{pc}) we get
\begin{equation}
p_{c}=-\left( 1+\omega \right) \rho \frac{\Gamma }{3H},  \label{A10}
\end{equation}%
and replacing in (\ref{A9}) we get
\begin{equation}
p_{c}=\frac{2\pi }{\sigma N}\left( 1+\omega \right) \left( 1+q\right), 
\label{eq: pc2}
\end{equation}%
where we have used (\ref{friedman}). This is the explicit expression that the \textit{creation} pressure takes in our model. Notice that if we assume $\Gamma >0$, which implies as we have mentioned before that $q<-1$, this necessarily means that $p_c<0$, under the consideration of a cold dark matter or quintessence EoS parameter $\omega$. We discuss a explicit examples in the next section.

\section{The cosmology}

Here we study the cosmology that emerges from the $\Gamma$ found in the previous section. Assuming a flat FLRW metric the equations of the model are
\begin{eqnarray}
3H^{2} &=&\rho ,  \label{A11} \\
2\dot{H}+3H^{2} &=&-\left( p+p_{c}\right) ;  \label{A12} \\
\dot{\rho}+3H\left( \rho +p+p_{c}\right) &=&0,  \label{A13}
\end{eqnarray}%
where $\rho $ is the energy density of the created matter and $p_{c}$ is given by (\ref{eq: pc2}). This system of equations considers a {\it single} component characterized by the energy density $\rho$ and the pressure $p$ where the particle number is changing by a process of particle creation (if $\Gamma$ is positive).

By combining (\ref{A12}) with the expression for \(p_c\) derived in (\ref{eq: pc2}), we arrive at the following key relation:
\begin{equation} \label{eq: keyrel}
    -\frac{2}{3}(1+q) + (1+\omega) = -\frac{2\pi}{\sigma N \rho} (1+\omega)(1+q).
\end{equation}
This equation reveals a striking feature: under the assumption \(q < -1\), consistent with a matter-creation-driven expansion, the relation remains satisfactorily consistent for \(\omega > -1\). This constraint leads naturally to a quintessence-like regime or, more intriguingly, a non-relativistic matter component. The latter is of particular interest as it circumvents the need for introducing exotic components, offering a straightforward yet profound explanation within the framework.

In the non-relativistic case, the above equation simplifies and directly leads to:
\begin{equation}\label{eq: qrel}
    1+q = - \frac{3N\sigma \rho}{6\pi - 2 N \sigma \rho} < 0.
\end{equation}
This result is remarkable, as it confirms consistency with \(q < -1\), providing robust theoretical support for the proposed phase of matter creation. The implications of this mechanism extend beyond the avoidance of exotic components, positioning it as a compelling alternative for describing cosmic evolution.

Another way to see the main conclusion of this section, is by using an effective description. Let us write (\ref{A13}) in the form
\begin{equation}
    \dot{\rho} + 3H(1+\omega_{\text{eff}})\rho = 0,
\end{equation}
where we have defined $\omega_{\text{eff}}= \omega + p_c/\rho$. After replacing the value we have found for $p_c$ we find
\begin{equation}
    \omega_{\text{eff}} = \omega - \frac{2\pi}{3\sigma N\rho} (1+\omega)(|q|-1).
\end{equation}
Let us first consider the case for a pure non-relativistic component with $\omega=0$. In this case,
\begin{equation}
\omega_{\text{eff}} = -\frac{2\pi}{3\sigma N\rho} \left( |q| - 1 \right). \label{A16}
\end{equation}
we notice that $\omega_{\text{eff}}$ can be in the range of quintessence or phantom regime.
These expressions emphasize the role of $\sigma N \rho$ in determining the regime of $\omega_{\text{eff}}$. Therefore, if we require $\Gamma > 0$, it follows that $q < -1$, which implies $p_c < 0$ if $\omega > -1$ (corresponding to cold dark matter or quintessence). This subtle interplay of parameters constitutes the core contribution, or the essence, of the present work.

From this, the \textit{phantom} condition imposes the constraint \(\sigma N\rho < 3\pi\). Since \(\sigma N\rho\) is clearly positive, there is little more to elaborate on at this stage. \textit{Future work} will aim to investigate \(\sigma N\rho\) in greater depth.

\section{Conclusions}

In this work, we have developed a novel framework for matter creation cosmology, beginning with a single-component fluid characterized by a constant equation of state parameter, \(\omega\). By introducing a generalized second law of thermodynamics and incorporating the entropy associated with the cosmic horizon, we established a profound connection between thermodynamics and cosmology. Crucially, by demanding that the cosmic expansion be adiabatic, we uniquely fixed the particle creation rate \(\Gamma\)—a result that, to the best of our knowledge, has no precedent in the existing literature on matter creation models.

This approach leads to a remarkable cosmological scenario. The model naturally produces a phantom-like expansion evolution while relying solely on a single constituent. Depending on the choice of \(\omega\), this constituent can range from a quintessence-like fluid to a completely non-exotic, non-relativistic dark matter fluid. This avoids the need for introducing any exotic physics, offering a compelling and elegant alternative to describe the accelerated expansion of the universe. 

The implications of this work extend beyond theoretical appeal. By unifying matter creation dynamics with thermodynamic principles, we provide a robust and consistent framework that opens new avenues for exploring the interplay between cosmic expansion, horizon thermodynamics, and particle creation. This novel mechanism not only challenges the conventional understanding of phantom cosmologies but also highlights the richness and potential of thermodynamic considerations in shaping the evolution of the universe.

We anticipate that this work will inspire further exploration of adiabatic particle creation mechanisms, their observational signatures, and their broader implications in modern cosmology.

Imposing adiabaticity within the particle creation framework (\(\Gamma > 0\)) inherently leads to the emergence of a \textit{phantom} regime. The theoretical connection \(\Gamma > 0 \longleftrightarrow q < -1\) and \(p_c < 0\) forms the central focus of this work. It is also worth noting that the presence of a \textit{phantom} regime is intrinsically tied to spacetime singularities. Investigating the nature of these singularities represents an intriguing direction for future research, with the potential to deepen our understanding of \textit{phantom} dynamics and their cosmological implications.

\section*{Acknowledgments}
VHC would like to thank CEFITEV-UV for partial support.


\end{document}